\documentclass[twocolumn,showpacs,preprintnumbers,amsmath,amssymb]{revtex4}
\usepackage{graphicx}
\usepackage{dcolumn}
\usepackage{bm}
\usepackage{multirow}

\begin{document}

\title{Spectral properties of photon pairs generated by spontaneous four wave mixing in inhomogeneous photonic crystal fibers}

\author{Liang Cui, Xiaoying Li}
\email[Corresponding author:~]{xiaoyingli@tju.edu.cn}
\author{Ningbo Zhao}
\affiliation{College of Precision Instrument and Opto-electronics Engineering, Tianjin University, Key Laboratory of Optoelectronics Information Technology of Ministry of Education, Tianjin 300072, P. R. China}

\begin{abstract}

The photonic crystal fiber (PCF) is one of the excellent media for generating photon pairs via spontaneous four wave mixing. Here we study how the inhomogeneity of PCFs affect the spectral properties of photon pairs from both the theoretical and experimental aspects. The theoretical model shows that the photon pairs born in different place of the inhomogeneous PCF are coherently superposed, and a modulation in the broadened spectrum of phase matching function will appear, which prevents the realization of spectral factorable photon pairs. In particular, the inhomogeneity induced modulation can be examined by measuring the spectrum of individual signal or idler field when the asymmetric group velocity matching is approximately fulfilled. Our experiments are performed by tailoring the spectrum of pulsed pump to satisfy the specified phase matching condition. The observed spectra of individual signal photons, which are produced from different segments of the 1.9 m inhomogeneous PCF, agree with the theoretical predictions. The investigations are not only useful for fiber based quantum state engineering, but also provide a dependable method to test the homogeneity of PCF.

\end{abstract}


\maketitle

\section{introduction}

\noindent
Spontaneous four wave mixing (SFWM) is a parametric process in which two
pump photons at frequency $\omega _{p1}$ and $\omega _{p2}$ scatter through the $\chi ^{(3)}$ (Kerr)
nonlinearity to create
energy-time entangled signal and idler photons at frequencies $\omega _s$ and $
\omega _i$, respectively, such that $\omega _{p1}+\omega _{p2}=\omega _s+\omega
_i$. To enhance the probability of SFWM, the phase matching condition determined by the dispersion of nonlinear medium needs to be satisfied. Among the various kinds of Kerr nonlinear materials, such as optical fibers, silicon wave-guides, and atomic media etc.~\cite{LJWang-2001,Fiorentino2002,Sharping2006-waveguide,Harris-PRL2005},
the highly nonlinear photonic crystal fiber (PCF) is identified as a promising candidate for developing the practical sources of photon pairs with low cost and compact size, owing to its advantages of controllable dispersion and excellent spatial mode purity~\cite{Sharping04PCF,Rarity2005,Fan2005,Garay-Palmett-OE2007}.

For the photon pairs generated by pumping the PCF with a pulsed laser,
the length of PCF, usually from tens of centimeters to a few meters, is not only related to the brightness, but also related to the spectral factorability~\cite{Rarity2005,Fan2005,Garay-Palmett-OE2007,Walmsley-tailored,Rarity-TAILORED,C-Silberhorn-TAILORED}. For example, to produce asymmetrical spectral factorable photon pairs, which is important for the conditional-preparation of pure single photons~\cite{URen05LP}, in addition to the satisfaction of asymmetric group velocity matching (AGVM) condition, the PCF with a homogeneous structure and with a longer length is desirable. However, unavoidable fluctuations in
PCF manufacturing process may have a large effect on their dispersion due to the small core
size and strong dispersion dependence on the spatial structure. Therefore, the factorability of photon pairs is restricted by the limited maximum length of PCF~\cite{Walmsley-tailored,Rarity-TAILORED}. The problem has been known for some time, but the detailed investigation about this issue has not been done yet.

In this paper, we study how the inhomogeneity of PCFs affect the spectral properties of photon pairs from both the theoretical and experimental aspects. Our theoretical model shows that the photon pairs born in different place of the inhomogeneous PCF are coherently superposed, and a modulation in the broadened spectrum of phase matching function will appear, which prevents the realization of spectral factorable photon pairs. In particular, the inhomogeneity induced modulation can be examined by measuring the spectrum of individual signal or idler field when the AGVM condition is approximately fulfilled. Our experiments are performed by tailoring the spectrum of pulsed pump to approximately satisfy the AGVM condition, and the observed spectra of individual signal photons, which are produced from different segments of the 1.9 m inhomogeneous PCF, agree with the theoretical predictions. Moreover, our results indicate that in the sense of generating spectral factorable photon pairs, the problem of lacking the homogeneous PCF with a longer length might be overcome in some extent by selecting and splicing the shorter PCFs having almost identical dispersion properties. The investigations are not only useful for fiber based quantum state engineering, but also provide a dependable method to test the homogeneity of PCF.

\section{Theoretical analysis}
Using a simplified model of the inhomogeneous PCF, we first analyze the joint spectral amplitude (JSA) of photon pairs, and propose to investigate the influence of nonuniform dispersion on phase matching function by examining the spectrum of individual signal or idler field under the AGVM condition. Then we simulate the spectral properties of individual signal photons in specified inhomogeneous PCFs when the AGVM condition is approximately satisfied. The simulations show that the inhomogeneity will induce a modulated pattern in the broadened spectrum of the phase matching function, which results in a reduced factorability of JSA.


\subsection{Phase matching function of spontaneous four wave mixing in inhomogeneous PCF}

In the theoretical model, we assume the signal and idler fields generated via the pulsed pumped SFWM in PCF are linearly co-polarized with the pump, and detuning between signal and idler photons is large enough to get rid of the contamination of Raman scattering~\cite{Rarity2005,Fan2005}. Moreover, the bandwidth of individual signal (idler) field is naturally narrow band, so the generation of photon pairs is basically filter free~\cite{URen05LP,Garay-Palmett-OE2007}.

The Gaussian shaped transform limited pump pulses remain classical, and the pump field propagating along the PCF is expressed as
\begin{equation}\label{fieldp}
E_p^{(+)}=E_{p0}e^{ - i\gamma P_p z}\int d\omega _p \exp[-\frac{(\omega _p-\omega _{pc})^2}{2\sigma _p^2}]e^{i(k_pz-\omega _pt)},
\end{equation}
where $\gamma$ is the nonlinear coefficient of PCF, $P_p$ is the peak pump power, $\omega _p$ is the frequency of pump photons, $\omega_{pc}$ and $\sigma_p$ are the central frequency and bandwidth of the pump, respectively, and $E_{p0}$ is associated with the peak pump power through $P_p\propto E_{p0}^2\sigma_p^2$. In the low gain regime, the field operator of the signal (idler) photons at the output of PCF is given by~\cite{Alibart06,Ma11}
\begin{equation}\label{operator}
a(\omega_{s(i)})=a_0(\omega_{s(i)})+\frac{G}{\sigma_p}\int d\omega_{i(s)}f(\omega_s,\omega_i)a_0 ^{\dag}(\omega_{i(s)})+o(G),
\end{equation}
where $a_0(\omega_{s(i)})$ and $a_0 ^{\dag}(\omega_{s(i)})$ are the annihilation and creation operators of the vacuum fields at $\omega_{s}$ ($\omega_{i}$), respectively, and the parameter $G\propto \gamma P_p $ is proportional to the gain of SFWM. The JSA function $f(\omega_s,\omega_i)$ can be expressed as the product of the pump envelop and the phase matching function:
\begin{equation}\label{jsa}
f(\omega_s,\omega_i)=\alpha(\omega_s,\omega_i)\times\phi(\omega_s,\omega_i).
\end{equation}
Accordingly, the joint spectral intensity (JSI) is written as
\begin{equation}
F(\omega_s,\omega_i)=|f(\omega_s,\omega_i)|^2,
\end{equation}
which is proportional to the probability of one pair of signal and idler photons emerging at $\omega_s$ and $\omega_i$, respectively.

The pump envelope can be well determined. When the self phase modulation induced spectral broadening of the pump is negligible, the pump envelop function can be approximated as~\cite{Alibart06,yang11}
\begin{equation}\label{pump-envelop}
\alpha(\omega_s,\omega_i)=\exp[-\frac{(\omega_s+\omega_i-2\omega_{pc})^2}{4\sigma_p^2}].
\end{equation}
While the phase matching function is written as
\begin{equation}\label{phi}
\phi(\omega_s,\omega_i)=\int_0^L \exp(i\Delta k z) dz,
\end{equation}
where $L$ is the total length of PCF, $\Delta k=2k(\omega_p)-k(\omega_s)-k(\omega_i)-2\gamma P_p$ is the wave vector mismatch, in which $k(\omega_j)$ ($j=p,s,i$) refers to the wave vector at frequencies $\omega_j$, and the term $2\gamma P_p$ can be neglected due to its smallness~\cite{Alibart06,Rarity2005}. Using the linear approximation of  $\Delta k$ by expanding the wave vectors $k(\omega_p)$, $k(\omega_s)$, and $k(\omega_i)$ at the perfect phase matching frequencies, $\omega_{pc}$, $\omega_{s0}$ and $\omega_{i0}$, respectively, we arrive at
\begin{equation}\label{delta-k}
\Delta k=\tau_s(\omega_{s}-\omega_{s0})+\tau_i(\omega_{i}-\omega_{i0}),
\end{equation}
where $\tau_{s(i)}=k'_{pc}-k'_{s(i)0}$, with $k'_l=dk(\omega)/d\omega|_{\omega=\omega_j}$ ($l=pc,s0,i0$) representing the reciprocal of group velocity, is the group velocity mismatch between the pump and signal (idler) fields.

From Eqs. (3), (5) and (6), one sees that the key to figure out the characteristics of JSA is the phase matching function.
If the PCF is homogeneous in structure, the wave vector $k(\omega_j)$ ($j=p, s, i$) and the mismatch term $\Delta k$ are constant along the fiber. By carrying out the integration in Eq. (\ref{phi}), we obtain the phase matching function \begin{equation}
\phi(\omega_s,\omega_i)=L\mathrm{sinc}(\frac{\Delta kL}{2})\exp(i\frac{\Delta kL}{2}).
\end{equation}

\begin{figure}[htpb]
\centering\includegraphics[width=8cm]{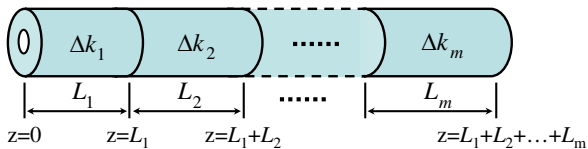}
\caption{(Color online) Schematic illustration of an inhomogeneous PCF consisting of
$m$ segments. Each segment is homogeneous, but the dispersion and length of different segments deviate.}
\label{model}
\end{figure}

If the PCF is inhomogeneous in structure, on the other hand, $k(\omega_j)$ ($j=p, s, i$) and $\Delta k$ will change along the fiber. In this situation, let us analyze the phase matching function by dividing the PCF into $m$ segments, as shown in Fig. \ref{model}. For the sake of simplicity, we assume each segment is homogeneous, but the dispersion and length of different segments deviate. For the $n$th segment, the length is $L_n$ and the wave vector mismatch is $\Delta k_n$. By performing a piecewise integration in Eq. (\ref{phi}), the overall phase matching function for the assembly of $m$ segments can be calculated and expressed as
\begin{equation}
\begin{aligned}\label{phi-inhomogeneous}
&\phi(\omega_s,\omega_i)=L_1\mathrm{sinc}(\frac{\Delta k_1L_1}{2})\exp(i\frac{\Delta k_1L_1}{2})\\
&+\sum\limits_{n=2}^m L_n\mathrm{sinc}(\frac{\Delta k_nL_n}{2})\exp(i\frac{\Delta k_nL_n}{2})\exp(i\sum\limits_{l=1}^{n-1} \Delta k_lL_l),
\end{aligned}
\end{equation}
which shows that photon pairs born in different place of the PCF is coherently combined.
In principle, it is straightforward to visualize the spectral property of $\phi(\omega_s,\omega_i)$ by implementing a joint spectral coincidence measurement through a two-dimension scanning in the signal and idler bands to reconstruct the JSI function~\cite{C-Silberhorn-TAILORED}. However, this method is complicated and time consuming.

To acquire the spectral property of the phase matching function in a easier way, we further simplify the model by assuming that the AGVM condition is approximately satisfied in each segment of the inhomogeneous PCF, i.e. $\tau_i\approx0$ or $\tau_s\approx0$. Defining the angle $\theta=|\arctan(\tau_{i}/\tau_{s})|$ as the acute angle between the $\omega_i$ axis and the contour of $\phi(\omega_s,\omega_i)$ in the $\{\omega_s,\omega_i\}$ space, we then have $\theta=0$ and $\theta=\pi/2$ for the case of $\tau_i$=0 and $\tau_s$=0, respectively~\cite{Garay-Palmett-OE2007}. Under the AGVM condition of $\tau_{i(s)}\approx0$, the argument $\tau_{i(s)}$ vanishes in the expression of $\phi(\omega_s,\omega_i)$ (see Eqs. (6) and (7)). In this situation, the projection of JSI on the signal (idler) axis, i.e., the spectrum of individual signal (idler) photons, is
\begin{equation}
\begin{aligned}\label{projection}
F_{s(i)}(\omega_{s(i)})&=\int|\alpha(\omega_s,\omega_i)\times\phi(\omega_{s(i)})|^2 d\omega_{i(s)}\\
&=\sqrt{2\pi}\sigma_p |\phi(\omega_{s(i)})|^2 \propto |\phi(\omega_{s(i)})|^2,
\end{aligned}
\end{equation}
which is irrelevant to the envelop of pump even if the spectral broadening of pump pulses becomes observable.

Equation (10) indicates that we can examine the phase matching function in the one dimensional space by experimentally measuring the spectrum of the individual signal (idler) photons. For example,
in the case of $\tau_i\approx0$, the spectrum of $|\phi(\omega_{s})|^2$ can be deduced by applying a tunable narrow band filter with a bandwidth of $\sigma_s$ in signal field. When the central frequency of the Gaussian shaped filter $\omega'_s$ is scanned, the photon counting rate of sampled signal photons (per pulse) can be expressed as
\begin{equation}
\begin{aligned}\label{projection2}
S_{s}(\omega'_{s})\propto\frac{|G|^2}{\sigma_p} \int |\phi(\omega_{s})|^2 \exp [-\frac{(\omega_s-\omega'_s)^2}{\sigma_s^2}]d\omega_{s},
\end{aligned}
\end{equation}
which shows that $|\phi(\omega_{s})|^2$ can be extracted from the measured $S_{s}(\omega'_{s})$. In practice, we prefer to use a filter with $\sigma_s$ much smaller than the bandwidth of $|\phi(\omega_{s})|^2$, so that the spectrum of $S_{s}(\omega'_{s})$ itself is close to that of $|\phi(\omega_{s})|^2$.

\subsection{Simulated spectrum of phase matching function in inhomogeneous PCFs}

In order to figure out the major spectral characteristics of the photon pairs,
we simulate the spectrum of the phase matching function under the AGVM condition.
In the simulations, the inhomogeneous PCFs are the random combinations of a few elemental segments, whose effective core radius $r$ and air-filling fraction $f$ are shown in Table \ref{table1}. For convenience, the four kinds of segments in Table \ref{table1} are labeled as S1, S2, S3 and S4, respectively. We note that the structure parameters $r$ and $f$ in Table \ref{table1} are selected by referencing the average effective values of the PCF used in our experiment, which are implied from the measured dispersion data of the PCF (see Fig. \ref{gvd}) rather than directly measured with an electron micrograph image. According to the values of $r$ and $f$ of each segment, we then calculate the dispersion and phase matching condition by using the step-index fiber model~\cite{WONG-PCF-GVD}. The results of the phase matched wavelengths of signal and idler fields, $\lambda_{s0}$ and $\lambda_{i0}$, group velocity mismatch $\tau_{s}$, and the angle $\theta$ for the pump with central wavelength $\lambda_{pc}=2\pi c/\omega_{pc}=1070$ nm are also listed in Table \ref{table1}. Comparing the parameters of each segment, we find that for a 0.1\% variation in the core radius $r$, the shift of $\lambda_{s0}$ is notable, but the change of $\theta$ is ignorable. The results indicates that the AGVM condition $\tau_i=0$ is approximately fulfilled in each segment for $\lambda_{pc}=1070$ nm, and the argument $\omega_i$ in the overall phase matching function of $\phi(\omega_s,\omega_i)$ (see Eq. (9)) can be neglected.


\begin{table}[htb]
\centering\caption{The structure parameters and phase matching conditions for $\lambda_{pc}=1070$ nm of the elemental segments}\label{table1}
\begin{tabular}[c]{ccccc}
\hline
Label of segment  &  S1 & S2 & S3 & S4\\
\hline
$r$ (nm) & 947 & 947.5 & 948 & 948.5\\
$f$ &  29.6\% & 29.6\%  & 29.6\% & 29.6\%\\
\hline
$\lambda_{s0}$ (nm) & 1409.9 & 1413.6& 1417.3 &1421\\
$\lambda_{i0}$ (nm) & 862.1 & 860.8& 859.4 &858.1\\
$\tau_{s}$ (ps/m) & 3.2 & 3.2 & 3.3 & 3.4\\
$\theta$ (rad) & $\sim$0.004 & $\sim$0.002 & $\sim$0.001 &  $\sim$0.004\\
\hline
\end{tabular}
\end{table}

\begin{figure}[htpb]
\centering\includegraphics[width=7cm]{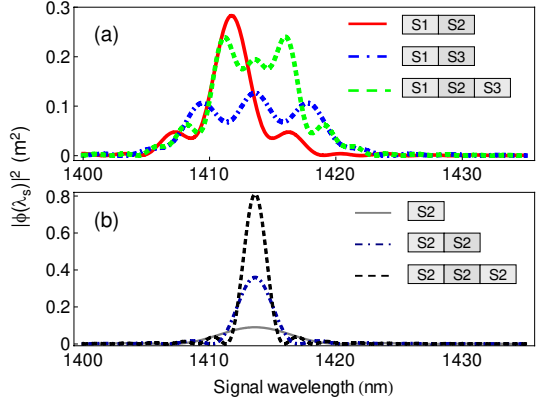}
\caption{(Color online) Simulated $|\phi(\lambda_s)|^2$ for (a) inhomogeneous PCFs with different configurations and (b) homogeneous PCFs with length of $L=0.3$ m, $L=0.6$ m, and $L=0.9$ m, respectively. For the inhomogeneous PCFs, the length of each segment is 0.3 m.}\label{sim-sig}
\end{figure}

Assuming the length of each elemental segment is 0.3 m, and substituting the parameters in Table \ref{table1} into Eqs. (\ref{delta-k}) and (\ref{phi-inhomogeneous}), we first calculate $|\phi(\lambda_{s})|^2$ for the inhomogeneous PCFs with the configuration of (i) S1+S2, (ii) S1+S3, and (iii) S1+S2+S3, respectively, as shown in Fig. \ref{sim-sig}(a). As a comparison, we also calculate $|\phi(\lambda_{s})|^2$ for the homogeneous PCFs with the spatial structure the same as S2, and with the total length $L$ of 0.3 m, 0.6 m, and 0.9 m, respectively, as shown in Fig. \ref{sim-sig}(b).  It is obvious that the spectral property of $|\phi(\lambda_{s})|^2$ in Fig. \ref{sim-sig}(a) is more complicated than that in Fig. \ref{sim-sig}(b). The modulated spectrum in Fig. \ref{sim-sig}(a) indicates that the counting rate of signal photons sampled by a narrow band filter centering at $\lambda'_s=2\pi c/\omega'_{s}$ may increase or decrease with the variation of $\lambda'_s$ (see Eq. (11)), and the variational range of counting rate can not be explained by the oscillations of a sinc-function. Moreover, the relative height of $|\phi(\lambda_{s})|^2$ in Figs. \ref{sim-sig}(a) and \ref{sim-sig}(b) show that for a given pump power, the intensity (per meter, per mode) of SFWM in homogeneous PCFs could be higher~\cite{Wong05JOSAB}.

\begin{figure}[htpb]
\centering\includegraphics[width=7.5cm]{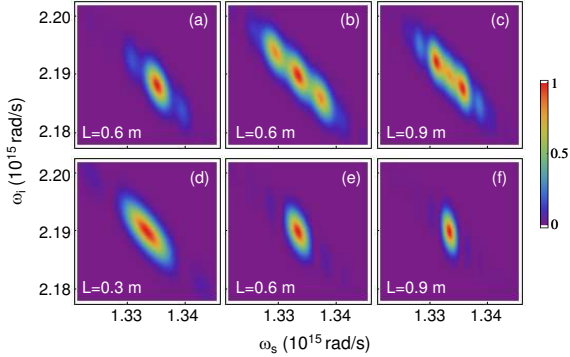}
\caption{(Color online) Normalized JSI $F(\omega_s,\omega_i)$ for inhomogeneous PCFs with the configuration of (a) S1+S2, (b) S1+S3, and (c) S1+S2+S3, respectively, and for homogeneous PCFs with the spatial structure the same as S2 and with the length of (d) $L=0.3$ m, (e) $L=0.6$ m, and (f) $L=0.9$ m, respectively.}\label{sim-jsi}
\end{figure}

For each PCF used in Fig. 2, we also plot the normalized JSI in the $\{\omega_s,\omega_i\}$ space, as shown in Fig. \ref{sim-jsi}. In the calculation, the full width at half maximum (FWHM) of pump, $\Delta \lambda_p$, is assumed to be 2 nm. Comparing Fig. \ref{sim-sig} with Fig. \ref{sim-jsi}, one sees that under the AGVM condition $\tau_i\approx0$, the spectrum of individual signal photons is a direct reflection of phase matching function (Eq. (10)). Moreover, comparing the plots of JSI for PCFs with the same length, it is obvious that the projections of JSI on both the signal and idler axes are broadened due to the inhomogeneity of PCFs.

\begin{figure}[htpb]
\centering\includegraphics[width=6.5cm]{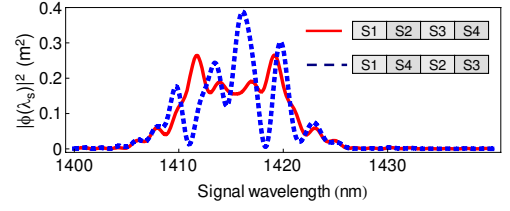}
\caption{(Color online) Simulated $|\phi(\lambda_s)|^2$ for inhomogeneous PCFs with the arranging order of the segments varied. The length of each segment is 0.3 m, and the total lengths of the PCFs are $L=1.2$ m. }\label{sim-sig3}
\end{figure}

Second, we calculate $|\phi(\lambda_{s})|^2$ when the arranging order of the elemental segments is altered. In the calculation, the PCFs consist of all the four kinds of segments listed in Table 1, the length of each segment is 0.3 m, but the order of the segments is arranged as (i)S1+S2+S3+S4 and (ii) S1+S4+S2+S3, respectively. As shown in Fig. \ref{sim-sig3}, the results of $|\phi(\lambda_{s})|^2$ for the two PCFs are obviously different, indicating the pattern of $|\phi(\lambda_{s})|^2$ can be varied by reordering the elemental segments.

\begin{figure}[htpb]
\centering\includegraphics[width=6.5cm]{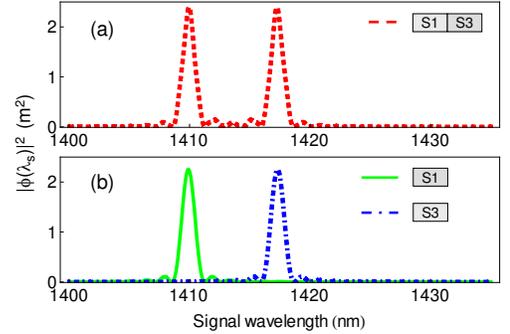}
\caption{(Color online) Simulated $|\phi(\lambda_s)|^2$ for (a) inhomogeneous PCFs consisting of two segments, and (b) 1.5 m homogeneous PCFs with the structures the same as S1 and S3, respectively. For the inhomogeneous PCF, the length of each segment is 1.5 m, and the total length is $L=3$ m.}\label{sim-sig2}
\end{figure}

Third, we compute $|\phi(\lambda_{s})|^2$ by changing the length of each segment to examine the  precondition for observing the modulated pattern, which can not be viewed as the oscillation of sinc-function. Figure \ref{sim-sig2}(a) shows the calculated $|\phi(\lambda_{s})|^2$ for the PCF with the configuration of S1+S3, and with the length of the each segment of 1.5 m. As a comparison, we also calculate $|\phi(\lambda_{s})|^2$ for the homogeneous PCFs with length $L=1.5$ m and with the structure the same as S1 and S3, respectively, as shown in Fig. \ref{sim-sig2}(b). One sees that the spectrum in Fig. \ref{sim-sig2}(a) can be viewed as a direct combination of the two spectra in Fig. \ref{sim-sig2}(b), whose overlapped region is almost negligible. The results indicate that if the spectra of signal field in different segments of the inhomogeneous PCF do not overlap, the modulation in the overall spectrum of $|\phi(\lambda_{s})|^2$ will not appear.

\subsection{Simulated intensity correlation function of signal field in inhomogeneous PCFs}

Having analyzed how the inhomogeneity of PCF affect the spectrum of phase matching function, let us briefly discuss its influence on the factorability of photon pairs by calculating the intensity correlation function of individual signal (idler) field $g^{(2)}$ (without heralding) ~\cite{Ma11}. In general, the better the factorability is, the higher $g^{(2)}$ will be. For photon pairs in spectral factorable state, the individual signal (idler) field with $g^{(2)}=2$ is in single mode thermal state, and its timing uncertainty is minimized; while for photon pairs with perfect spectral correlation, the individual field with $g^{(2)}$ approaching 1 is in multi-mode thermal state, and its timing uncertainty are dependent on the coherence time of the individual field and the pulse duration of pump.
For the specified case of $\tau_{i}\approx0$, the bandwidth (or coherence time) of individual signal photons is determined by that of phase matching function $\phi(\omega_s)$, and the timing uncertainty of signal photons is determined by the ratio between pump pulse duration ($1/\Delta \lambda_p$) and the coherence time of signal field. With the reduction of the timing uncertainty, the number of temporal mode contained in the signal field reduces, and the value of $g^{(2)}$ increases.

According to the field operator in Eq. (\ref{operator}),  $g^{(2)}$ of the individual signal photons can be written as ~\cite{christ2011NJP,Ma11}
\begin{equation}
\begin{aligned}\label{g2}
g^{(2)}&=\frac{\int d\omega_s d\omega_{s'} \langle \hat{a}^{\dag}(\omega_s) \hat{a}^{\dag}(\omega_s') a(\omega_s)  a(\omega_s')\rangle}{\int d\omega_s \langle \hat{a}^{\dag}(\omega_s) a(\omega_s)\rangle \int d\omega_{s'} \langle \hat{a}^{\dag}(\omega_s') a(\omega_s')\rangle}\\
&=1+\frac{\int d\omega_sd\omega'_s|\int d\omega_i f^*(\omega_s,\omega_i)f(\omega'_s,\omega_i)|^2}{|\int d\omega_sd\omega_i|f(\omega_s,\omega_i)|^2|^2},
\end{aligned}
\end{equation}
which shows $g^{(2)}\leq 2$ because of the Schwartz inequality, and the equality holds if and only if the JSA can be factorized into $f(\omega_s,\omega_i)=h(\omega_s)\times k(\omega_i)$. Here $h(\omega_s)$ and $k(\omega_i)$ refer to the probability amplitude of the signal and idler photons having the frequency $\omega _s$ and $\omega _i$, respectively. In general, in order to increase the value of $g^{(2)}$, one should properly control the energy and momentum conservation by regulating the pump pulse and by tailoring the dispersion of PCF~\cite{Garay-Palmett-OE2007}. For example, under the phase matching condition of $\tau_i\tau_s<0$, the bandwidth of pump envelope and that of phase matching function should be properly matched so that the frequency correlation terms in $\alpha(\omega_s,\omega_i)$ and  $\phi(\omega_s,\omega_i)$ can cancel out~\cite{URen05LP}, and the timing uncertainty of signal (idler) photons is minimized. However, for the inhomogeneous PCF, one sees that the corresponding JSA contains at least two addends due to nonuniform dispersion (see Eqs. (3), (5), and (9)), which prevents the formation of a factorable JSA.

For the AGVM condition of $\tau_{i}\approx0$, we compute $g^{(2)}$ by substituting the parameters of each PCF used in Figs. \ref{sim-sig}$-$\ref{sim-sig2} into Eqs. (\ref{jsa}), (\ref{pump-envelop}), (\ref{phi-inhomogeneous}), and (\ref{g2}). In the computation, the FWHM of pump $\Delta \lambda_p$ is assumed to be 2 and 5 nm, respectively. As shown in Table \ref{table2}, it is clear that for each kind of PCF, $g^{(2)}$ increase with $\Delta \lambda_p$ due to the accordingly reduced time uncertainty of signal photons.

\renewcommand{\multirowsetup}{\centering}
\begin{table}[htb]
\centering\caption{Calculated $g^{(2)}$ of individual signal field for $\lambda_{pc}=1070$ nm}\label{table2}
\begin{tabular}{c|c|c|c|c}
\hline
\multirow{3}{*}{PCF} & \multirow{3}{2.2cm}{Configuration} & \multirow{3}{1.8cm}{Total Length $L$ (m)} & \multicolumn{2}{c}{Calculated $g^{(2)}$} \\\cline{4-5}
 & & & $\Delta \lambda_p=$ & $\Delta \lambda_p=$ \\
 & & & 2 nm & 5 nm\\
\hline

\multirow{6}{1.5cm}{Inhomo- geneous} &S1+S2& 0.6&1.62&1.86\\\cline{2-5}

 &S1+S3&0.6 &1.43 & 1.75 \\\cline{2-5}

 & S1+S2+S3 & 0.9 & 1.52 & 1.82 \\\cline{2-5}

 & S1+S2+S3+S4 & 1.2 & 1.42 & 1.75 \\\cline{2-5}

 & S1+S4+S2+S3 & 1.2 & 1.44 & 1.76 \\\cline{2-5}

 & S1+S3 & 3 & 1.49 & 1.79 \\\cline{2-5}

\hline

\multirow{4}{1.5cm}{Homo- geneous} & \multirow{4}{*}{---} &0.3& 1.56 & 1.81 \\\cline{3-3} \cline{4-5}

 & &0.6& 1.75 & 1.90 \\\cline{3-3} \cline{4-5}

 & &0.9& 1.83 & 1.93 \\\cline{3-3} \cline{4-5}

 & &1.5& 1.89 & 1.96 \\

\hline
\end{tabular}
\end{table}

Now, let us analyze the data in Table \ref{table2} for the pump with a certain bandwidth. Obviously, $g^{(2)}$ obtained from an inhomogeneous PCF is less than that from a homogeneous PCF if the lengths of the two kinds of PCFs are equal. Moreover, in contrast to the homogeneous PCF, for which $g^{(2)}$ increases with the fiber length $L$, the relation between $g^{(2)}$ and $L$ for the inhomogeneous PCFs is not so straightforward. For example, for the 0.6 m inhomogeneous PCF with the configuration of S1+S2 and 0.9 m inhomogeneous PCF with the configuration of S1+S2+S3, respectively, the shorter one  corresponds to a greater $g^{(2)}$. Furthermore,
to figure out a common feature which closely related to $g^{(2)}$, we compare $g^{(2)}$ and the spectrum (see Fig. \ref{sim-sig} and Fig. \ref{sim-sig3}) for each inhomogeneous PCF. The results show that the key to obtain $g^{(2)}$ with a higher value is to narrow the overall spectrum of individual signal photons $F_{s}(\omega_{s})$ which is not only related to the length $L$, but also depends on dispersion property along the PCF. By comparing $g^{(2)}$ of inhomogeneous PCFs with the configuration of S1+S2, and S1+S3, respectively, one sees that the total lengths of the two PCFs are equal, but $g^{(2)}$ in the former case is higher. The results indicate that for the case of $\tau_{i}\approx0$, if the dispersion property of each segment is closer to each other, the overall spectrum of signal photon will be narrower, and the value of $g^{(2)}$ will be higher. Therefore, in the sense of generating spectral factorable photon pairs, the problem of lacking the homogeneous PCF with a longer length might be overcome in some extent by selecting and splicing the shorter PCFs having almost identical dispersion properties.

\section{Experiment}

\begin{figure}[htpb]
\centering\includegraphics[width=7cm]{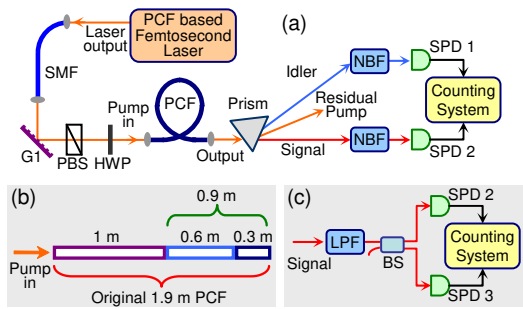}
\caption{(Color online) (a) Experimental setup, see the text for details. (b) Schematic of the cutting procedure of 1.9 m PCF. (c) Experimental setup for measuring the intensity correlation function $g^{(2)}$ of signal photons. }\label{setup}
\end{figure}

Our experimental setup is shown in Fig. \ref{setup}(a), signal and idler photon pairs are produced from the PCF (NL-1050-ZERO-2, Crystal Fibre), which is quoted to have a extremely flat, near-zero dispersion between 1025 nm and 1075 nm, the nonlinear coefficient $\gamma$ is $37$ W$^{-1}$km$^{-1}$ at 1064 nm, and the birefringence ($\Delta n$) is measured to be $2.5\times 10^{-5}$. The original length of the PCF is about 1.9 m (see Fig. \ref{setup}(b)). A compact mode-locked femtosecond Yb-doped fiber laser serves as the pump source~\cite{Song-PCF-laser}, whose linearly polarized output is centering at 1042nm, and FWHM is about 8 nm. The average power and repetition rate of the laser are about 400 mW and 54.4 MHz, respectively .

\begin{figure}[htpb]
\centering\includegraphics[width=6cm]{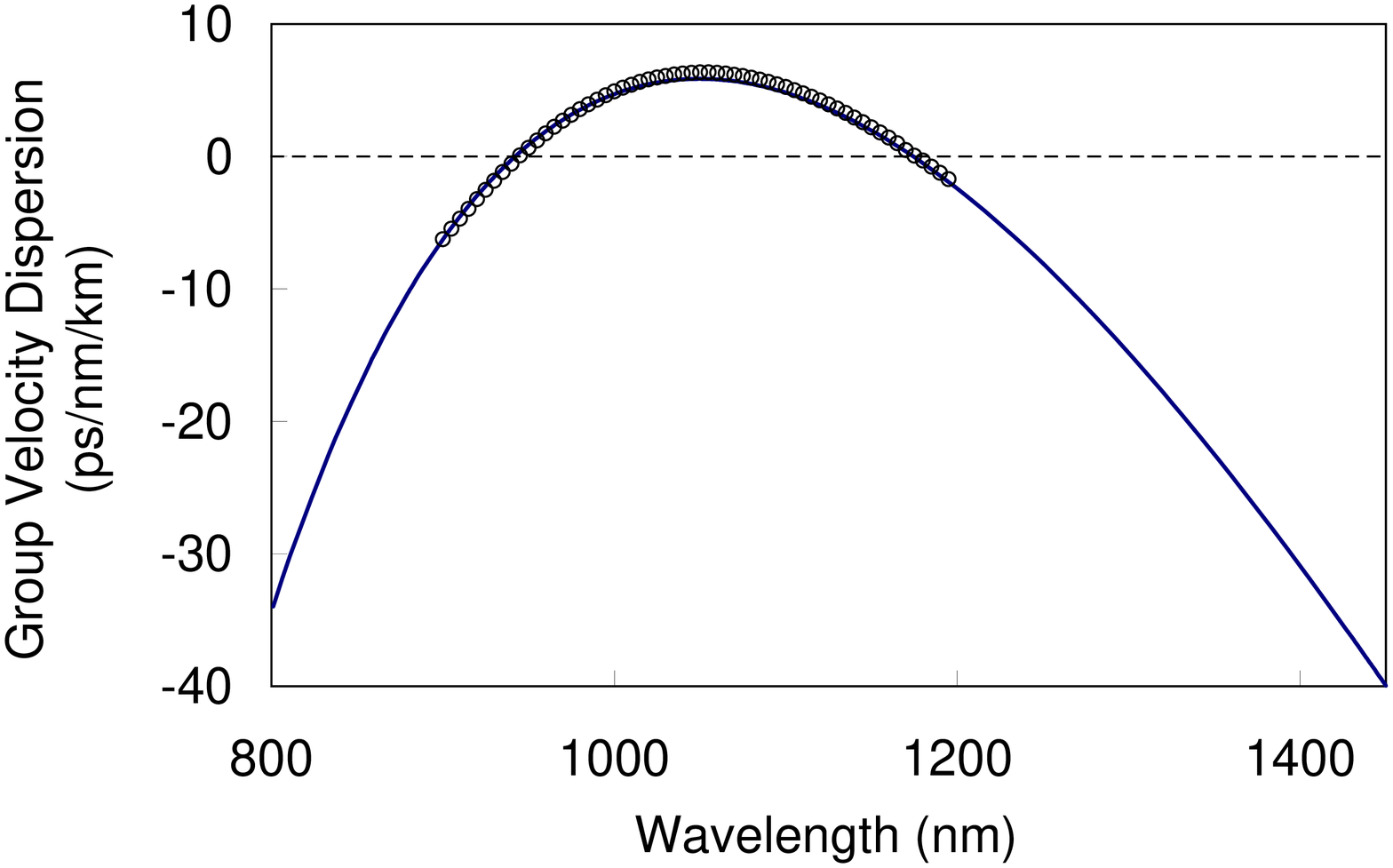}
\caption{(Color online) The group velocity dispersion measured by using the white light interferometry method. The hollow circles with the size representing the error bars are the data points, and the solid fitting curve is obtained by using the step-index model with $r=948$ nm and $f=29.6$\%.
}\label{gvd}
\end{figure}

\begin{figure}[htpb]
\centering\includegraphics[width=6cm]{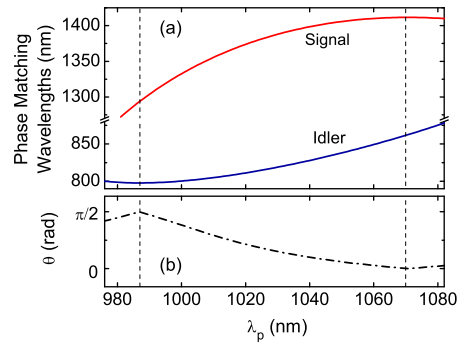}
\caption{(Color online) Calculated phase matching wavelengths and angle $\theta$ as a function of the pump wavelength $\lambda_{p}$. The vertical dashed lines correspond to $\lambda_{p}$=1070.2 nm and $\lambda_{p}$=986.5 nm, at which the AGVM conditions $\theta=0$ and $\theta=\pi/2$ ($\tau_i$=0 and $\tau_s$=0) are respectively satisfied.}\label{phase matching}
\end{figure}

We first characterize the average dispersion property of the 1.9 m PCF by utilizing the white light interferometry method. The broadband supercontinuum light is obtained by sending the output of the laser into a high nonlinear single mode fiber (SMF) and the group velocity dispersion of the PCF along its fast axis is measured in the wavelength range of 900 to 1200 nm. The result in Fig. \ref{gvd} indicates the two zero dispersion wavelengths are about $942 \pm4$ nm and $1175\pm4$ nm, respectively. Fitting the measured data by employing the step-index fiber model, we obtain the average effective core radius and air-filling fraction in the fast axis of the PCF are about $948\pm2$ nm and $(29.6\pm0.1)$\%, respectively. Moreover, we calculate the SFWM phase matching wavelengths and angle $\theta$ as a function of the pump wavelength $\lambda_{p}$, as shown in Figs. \ref{phase matching}(a) and \ref{phase matching}(b), respectively. One sees that the AGVM conditions, $\tau_i$=0 and $\tau_s$=0, corresponding to $\theta=0$ and $\theta=\pi/2$, respectively, are fulfilled for $\lambda_{p} \approx 1070.2$ nm and $\lambda_{p} \approx 86.5$ nm. Moreover, we find that in the vicinity of $\lambda_{p} \approx 1070.2$ nm and $\lambda_{p} \approx 986.5$ nm, respectively, the wavelength range of $\lambda_{p}$ satisfying $\theta\approx0$ is greater than that satisfying $\theta\approx\pi/2$. Considering the uncertainty of dispersion measurement, we set the central wavelength of pulsed pump $\lambda_{pc}$ at 1070 nm to guarantee the AGVM condition $\tau_i$=0 is approximately satisfied in the PCF.

To ensure the power of pulsed pump with $\lambda_{pc}$=1070 nm is sufficient to conduct our experiments, we replace the high nonlinear SMF with a 40 cm long regular SMF to expand the laser spectrum to 70 nm via self phase modulation effect, and use the grating G1 to carve out the required pump (see Fig. \ref{setup}(a)). The polarization of pump propagating through the polarization beam splitter (PBS) is adjusted by the half-wave-plate (HWP). The average pump power launched into the PCF is about 0.9 mW, the FWHM and pulse duration of the pump are about 2 nm and 0.8 ps, respectively.

To reliably detect the signal and idler photons, we need to reject the pump and to separate the signal and idler photons. We achieve this by placing a prism at the output port of PCF. Moreover, to resolve the spectrum of signal and idler photons, tunable narrow band-pass filters (NBFs) realized by gratings are used in both signal and idler bands. In the idler band, the FWHM of NBF is 0.4 nm; while in signal band, the FWHM can be adjusted to be either 3.5 nm or 0.8 nm. The idler photons are detected by silicon based single photon detector (SPD) SPD1 (SPCM-AQRH-15), and the signal photons are detected by InGaAs/InP based SPD2 (id200). For the counting measurement of individual signal photons, the 2.5-ns-wide gate pulses of SPD2 arrive at a rate of about $3.4$ MHz, $1/16$ of the repetition rate of the pump pulses; while for measuring the coincidence between signal and idler photons, the gate pulses are triggered by the detection events of SPD1. The total detection efficiencies, including the filtering system and the SPDs, are $\sim$4\% and $\sim$5\% in the signal and idler bands, respectively.

\begin{figure}[htpb]
\centering\includegraphics[width=5cm]{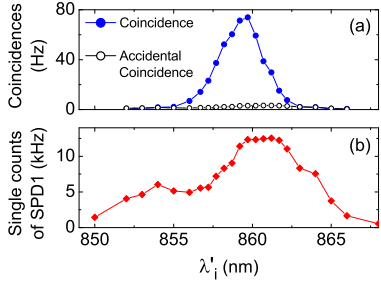}
\caption{(Color online) (a) Coincidence and accidental coincidence rates between SPD1 and SPD2, and (b) single counts of SPD1 versus the central wavelength of NBF in idler band $\lambda'_{i}$.}\label{idler-coin}
\end{figure}

We then test the quantum correlation properties of signal and idler photon pairs in the 1.9 m PCF. According to the theoretical calculation, the FWHM of signal and idler photons in a 1.9 m PCF with homogeneous structure are calculated to be about 0.9 and 1.8 nm, respectively. Therefore, to make sure most of the signal photons can propagate through the filter, the FWHM of the NBF in signal band is adjusted to be about 3.5 nm, which is about four times broader than the estimated bandwidth. In this experiment, the pump is launched along the fast axis of PCF, and the central wavelength of the NBF in signal band is fixed at 1417 nm, at which the single count rate of SPD2 is considerable. When the central wavelength of NBF in idler band $\lambda'_{i}$ is scanned from 850 to 868 nm, at each wavelength point, we not only measure the single counts of SPD1, but also measure the coincidences and accidental coincidences of SPD1 and SPD2  by recording coincidence rates of signal and idler photons produced in the same pulse and adjacent pulses, respectively. Figure \ref{idler-coin}(a) shows the measured coincidence rates are much greater than the accidental coincidence rates, clearly demonstrating the quantum correlation of the signal and idler photons via SFWM. However, we find the FWHM of coincidences in is about 3 nm, much greater than that of the estimated value of about $1.8$ nm. Moreover, the spectrum of single counts in the idler band shown in Fig. \ref{idler-coin}(b) is even much broader than that of coincidences, indicating the phase matching function of SFWM in the PCF can not be simply described by a sinc-function. Since we have verified the satisfaction of $\tau_i\approx0$ by demonstrating the frequency de-correlation of the signal and idler photon pairs in Ref. \cite{LiFIO10}, we think the spectrum broadening in Figs. \ref{idler-coin}(a) and \ref{idler-coin}(b) are caused by the inhomogeneous dispersion along the PCF.

\begin{figure}[htpb]
\centering\includegraphics[width=7.5cm]{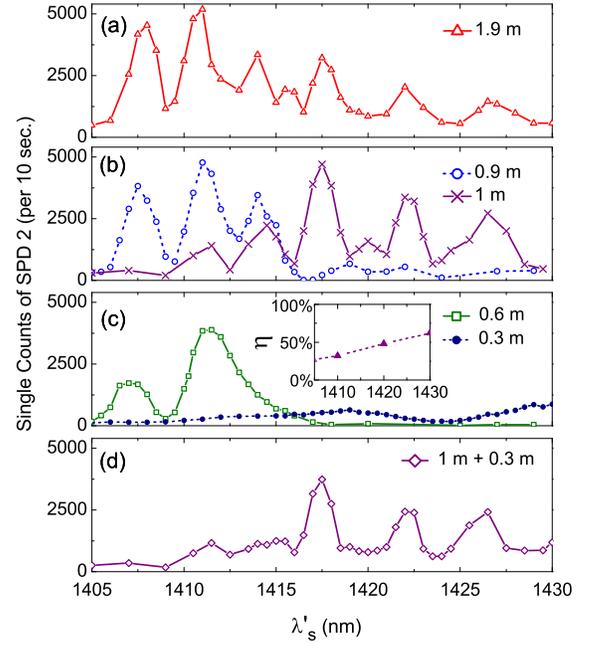}
\caption{(Color online) The recorded single counts of SPD2 as the central wavelength of NBF in signal band, $\lambda'_{s}$, is varied from 1405 to 1430 nm for (a) the original 1.9 m PCF, (b) the 0.9 m and 1 m PCFs, respectively, (c) the 0.6 m and 0.3 m PCFs, respectively, and (d) the rejoined 1.3 m PCF. The inset in plot (c), illustrating the transmission efficiency per meter, $\eta$, versus the wavelength of signal photons, is obtained by using the cutting back method. The curves connecting the data points are only for guiding the eyes. }\label{signal-spectra}
\end{figure}

Next, in order to confirm the 1.9 m PCF is inhomogeneous, we investigate the spectrum of the overall phase matching function by measuring the full spectrum of the individual signal photons (Eq. (11)). Since the filter with a narrower bandwidth helps to directly achieve a fine spectrum structure, we adjust the FWHM of NBF in the signal band to 0.8 nm. Figure \ref{signal-spectra}(a) is obtained by recording the single counts of SPD2 as $\lambda'_{s}$ is varied from 1405 to 1430 nm. One sees that the spectrum with a modulated pattern spreads in a range up to 25 nm, and the heights of the observed 7 peaks can not be explained by the oscillation of sinc-function.

To exhibit the inhomogeneity of PCF more clearly, we cut the 1.9 m PCF into two segments, whose lengths are 1 m and 0.9 m (see Fig. \ref{setup}(b)), respectively. For each segments, we also record the single counts of SPD2 by varying $\lambda'_{s}$. As shown in Fig. \ref{signal-spectra}(b), the spectra of the two segments are about respectively responsible for the different parts of the spectrum in Fig. \ref{signal-spectra}(a). By comparing the overlapped spectral region between each segment and the original 1.9 m PCF, one sees that for the 1 m PCF, the main peak of spectrum experiences an obvious rising (see Figs. \ref{signal-spectra}(a) and \ref{signal-spectra}(b)). We think this is because the transmission efficiency $\eta$ in the PCF is rather low in the wavelength range of about 1.4 $\mu$m, as evidenced by the inset in Fig. \ref{signal-spectra}(c).

Fig. \ref{signal-spectra} (b) not only clarify the dispersion along the original 1.9 m PCF is non-uniform, but also indicate that the two segments of PCFs are inhomogeneous. To better understand the inhomogeneity of PCF, we further cut the 0.9 m PCF into two subsegments (see Fig. \ref{setup}(b)), whose lengths are 0.3 m and 0.6 m, respectively. Again, for each subsegment, we measure the single counts of SPD2 by varying $\lambda'_{s}$. As shown in the main plot of Fig. \ref{signal-spectra}(c), the spectral patterns of the two subsegments are dramatically different from each other, and the spectrum of their direct combination is distinct from that of the 0.9 m PCF. Moreover, the spectra in Fig. \ref{signal-spectra}(c) indicate that the two subsegments are also inhomogeneous.

We note that in carrying out the above experiments, we have also measured the spectra of signal photons produced from different segments of the 1.9 m PCF by launching the pump along the slow axis of the PCF. The spectral pattern of each segment is very similar to that obtained by propagating the pump along the fast axis, except a upshift in wavelength of about 2 nm. Therefore, the birefringence of the PCF does not contribute to the modulated spectra, and the data in Figs. \ref{signal-spectra}(a)-\ref{signal-spectra}(c) reflects how the spectra of phase matching function are affected by the random distribution of dispersion in inhomogeneous PCF. Although the dispersion of each segment cutting from 1.9 m PCF is non-uniform, which is different from that in the model shown in Fig. \ref{model}, the analysis in Sec. II qualitatively explains the experimental results.

It is worth pointing out that the signal spectrum of the original 1.9 m PCF seems like a direct summation of that of the two PCFs with lengths of 0.9 m and 1 m, respectively. One may think the phenomena might be an indication of the effective interaction length of SFWM owing to the walk-off between the signal and pump fields~\cite{Alibart06}. However, our investigation shows that the signal and idler photons of a pair should be considered as a whole, and no walk-off between the photon pairs and pump are observed~\cite{note01}. According to the analysis in Sec. II B, we think the reason responsible for the data in Figs. \ref{signal-spectra}(a) and \ref{signal-spectra}(b) is that the overlapped region between the signal spectra of the 0.9 m and 1 m segments is relatively small. If the overlapped region between the two segments is greatly increased, the signal spectrum of the inhomogeneous PCF would be different from a direct combination of that of each segment. To further verify this point, we rejoin the 1 m and 0.3 m PCFs by placing the two fiber ends in intimate contact. The transmission loss of the junction is $\sim$1 dB and the axes of the two PCFs are not accurately aligned due to the technical limitation. After launching the pump along the fast axis of the 1 m PCF, we measure the single counts of signal photons of the rejoined PCF when $\lambda'_{s}$ is varied. As shown in Fig. \ref{signal-spectra}(d), despite the imperfect connection, one sees that the influence of the 0.3 m PCF on the rejoined 1.3 m PCF is still observable. Comparing with the spectrum for 1 m PCF in Fig. \ref{signal-spectra} (b), we find that the heights of the two small peaks setting in the left side of the highest peak at about 1417 nm are depressed, and the small peak sitting in the right side of the highest peak is missing. The results show that direct combination of the signal spectra of the 0.3 m and 1 m PCFs is different from the spectral pattern of the rejoined PCF.

Finally, we study the influence of inhomogeneity on the factorability of photon pairs by measuring $g^{(2)}$ of the individual signal field. To do so, we perform the measurement by exploiting the setup in Fig. \ref{setup}(c). In the measurement, the condition $\tau_i\approx0$ still holds. The NBF is replaced with a long pass filter (LPF) with a cutoff wavelength of 1200 nm, and the FWHM in the signal band is $\sim$ 40 nm, which is exclusively provided by the prism. After passing the signal field through a 50/50 beam splitter (BS), the outputs of BS are fed into the SPD2 and SPD3 (id200), respectively. The value of $g^{(2)}$ is the ratio between the measured coincidence and accidental coincidence rates of SPD2 and SPD3~\cite{christ2011NJP,Ma11}. For the PCFs with length of 1 m and 0.6 m, the measured $g^{(2)}$ are $1.27\pm0.02$ and $1.56\pm0.02$, respectively. Since the bandwidths of signal photons in both PCFs are much smaller than 40 nm, the measurement can be viewed as a filter free case. Moreover, comparing the results of $g^{(2)}$ with the corresponding spectra the PCFs in Fig. \ref{signal-spectra}, one sees that the experimental results are consistent with our theoretical analysis.

\section{Conclusion}

In conclusion, we have investigated the spectral properties of the photon pairs generated from SFWM in inhomogeneous PCFs from the theoretical and experimental aspects. The theoretical model shows that the photon pairs born in different place of the inhomogeneous PCF are coherently superposed, and a modulation in the broadened spectrum of phase matching function will appear, which prevents the realization of spectral factorable photon pairs. In particular, the inhomogeneity induced modulation can be examined by measuring the spectrum of individual signal or idler field when the AGVM condition is approximately fulfilled. The experiment is performed by tailoring the spectrum of pump to ensure the AGVM condition is approximately satisfied. The observed spectra of individual signal photons, which are produced from different segments of the PCF, demonstrate the 1.9 m PCF and its segments with length down to 0.3 m are inhomogeneous, and the experimental results qualitatively agree with the theoretical predictions. Moreover, our results indicate that in the sense of generating spectral factorable photon pairs, the problem of lacking the homogeneous PCF with a longer length might be overcome in some extent by selecting and splicing the shorter PCFs having almost identical dispersion properties.

Our investigations are not only useful for fiber based quantum state engineering, but also provide a dependable method to test the inhomogeneity of PCF. Although it has been demonstrated that the nonuniform dispersion property in optical fibers can be characterized by using parametric amplification, our method is different from those reported in Ref. \cite{Karlsson98JOSAB} and \cite{Inoue94OL}. This is because (i) instead of using the amplified signal wave via continuous wave pumped four wave mixing in high gain regime, we exploit the pulse pumped SFWM in the low gain regime; and (ii) in contrast to characterize the fiber with length in the order of hundreds meters or kilometers, we can examine the inhomogeneity of PCF with length less than 1 m.

\begin{acknowledgments}
This work was supported in part by the NSF of China
(No. 11074186), the State Key
Development Program for Basic Research of China (No. 2010CB923101), and 111 Project B07014.

\end{acknowledgments}


\begin{thebibliography}{23}
\bibitem[{not()}]{note01}
\bibinfo{journal}{The work is being prepared for submission}.
\end{thebibliography}

\end{document}